\documentclass{article}
\RequirePackage{filecontents}
\PassOptionsToPackage{dvipsnames}{xcolor}
\RequirePackage{comgipp}
\RequirePackage{amsfonts}
\RequirePackage{amsmath}
\RequirePackage{authblk}

\newcommand{\theReferenceText}{\fullcite{Schubotz2020}}
\usepackage[
backend=biber,
style=numeric,
firstinits=true,
url=false,
isbn=false,
safeinputenc,%
]{biblatex}
\usepackage{amsmath}
\usepackage{hyperref}
\title{Mathematical Formulae in Wikimedia Projects 2020}

\author[1,2]{Moritz Schubotz}
\author[1]{Andr\'{e} Greiner-Petter}
\author[1,3]{Norman Meuschke}
\author[2]{Olaf Teschke}
\author[1,3]{Bela Gipp}
\affil[1]{University of Wuppertal, Germany ({andre.greiner-petter@zbmath.org, \{last\}@uni-wuppertal.de})}
\affil[2]{FIZ-Karlsruhe, Germany (\{first.last\}@fiz-karlsruhe.de)}
\affil[3]{University of Konstanz, Germany (\{first.last\}@uni-konstanz.de)}

\begin{document}
  \maketitle
  \thispagestyle{firststyle}
  \begin{abstract}
    This poster summarizes our contributions to Wikimedia's processing pipeline for mathematical formulae.
We describe how we have supported the transition from rendering formulae as course-grained PNG images in 2001 to providing modern semantically enriched language-independent MathML formulae in 2020.
Additionally, we describe our plans to improve the accessibility and discoverability of mathematical knowledge in Wikimedia projects further.
  \end{abstract}
\section{Introduction}\label{sec:intro}
Mathematical formulae are an integral part of Wikipedia and other projects of the Wikimedia foundation\footnote{List of Wikimedia projects: \url{https://meta.wikimedia.org/wiki/Wikimedia_projects}}. 
The MediaWiki software is the technical backbone of many Wikimedia projects, including Wikipedia. 
Since 2003, wikitext -- the markup language of MediaWiki -- supports mathematical content~\cite{Schubotz2014}.
For example, MediaWiki converts the wikitext code \verb|<math>E=mc^2</math>| to the formula $E=mc^2$.
While the markup for mathematical formulae has remained stable since 2003, MediaWiki’s pipeline for processing wikitext to render formulae has changed significantly. 

Initially, MediaWiki used LaTeX to convert math tags in wikitext to PNG images. 
The rendering process was slow, the images did not integrate well into the text, were inaccessible to screen readers for visually impaired users, and scaled poorly for both small and high-resolution screens. 
To alleviate these problems, we started developing a new JavaScript-based rendering backend called Mathoid in 2013 \cite{Schubotz2014}. 
Mathoid invokes MathJax on the server-side to convert the LaTeX code to MathML and SVG output. The new rendering pipeline became available in production in 2016~\cite{Schubotz2016}.

Improving the rendering of mathematical formulae was only a first step towards our ultimate goal of enhancing the discoverability of mathematical knowledge.  
Working towards that goal, we developed a first math search engine prototype for Wikipedia~\cite{Schubotz2012} in 2012.
However, we found that classic, lexical search functionality for mathematical content has little practical benefit.
The NTCIR MathIR competitions~\cite{Aizawa2014,Schubotz2015}, which will continue at the CLEF conference 2020, have confirmed our experience. The competitions use Wikipedia as a dataset to evaluate mathematical information retrieval systems. The NTCIR results indicate that systems employing established information retrieval technology fail to add significant value for the average Wikipedia user~\cite{Schubotz2016b}. 
Deploying math search to Wikipedia requires a semantic understanding of formulae, which in turn necessitates semantic augmentation of formulae.    

To increase the availability of semantic information on mathematical formulae, we implemented the rendering of formulae in Wikidata -- the central structured knowledge base for Wikimedia projects. 
The new functionality greatly facilitates the donation of semantically annotated mathematical formulae for volunteers. 

The availability of semantic formula data in Wikidata has thus far enabled several research projects, e.g., on math question answering~\cite{Schubotz2018a}, semantic augmentation of mathematical content~\cite{Scharpf2018}, and mathematical information retrieval~\cite{Scharpf2019b}. However, the connection between formulae in Wikidata and Wikipedia had no immediate benefit for the average Wikipedia user until January 2020.

\section{Enhanced formulae in Wikipedia}\label{sec:method}
\begin{figure}[tp]
  \centering
  \includegraphics[width=.9\columnwidth]{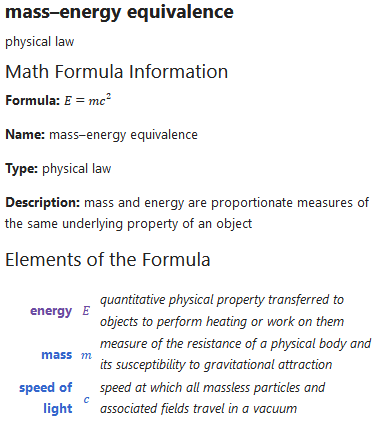}
  \caption{Semantic enhancement of the formula $E=mc^2.$ \small{\url{https://en.wikipedia.org/wiki/Special:MathWikibase?qid=Q35875}}}
  \providecommand{\Description}[2][]{}
  \Description[]{Special page on the mass-energy equivalence formula.}
  \label{figure}
  \vspace{-1.5em}
\end{figure}

In January 2020, we deployed a feature that enables enhancing mathematical formulae in Wikipedia with semantics from Wikidata.
For instance, the wikitext code \verb|<math qid=Q35875>E=mc^2</math>| now connects the formula $E=mc^2$ to the corresponding \href{https://www.wikidata.org/wiki/Q35875}{Wikidata item} by creating a hyperlink from the formula to the special page shown in Figure \ref{figure}. 
The special page displays the formulae together with its name, description, and type, which the page fetches from Wikidata.
This information is available for most formulae in all languages.
Moreover, the page displays elements of the formula modeled as \texttt{has part} annotations of the Wikidata item.

The \texttt{has part} annotation is not limited to individual identifiers but also applicable to complex terms, such as $\frac12m_0v^2$, i.e., the \href{https://en.wikipedia.org/w/index.php?oldid=939835125#Mass–velocity_relationship}{kinetic energy approximation for slow velocities}.
For example, we demonstrated using the annotation for the
\href{https://en.wikipedia.org/w/index.php?title=Special:MathWikibase&qid=Q1899432}{Grothendieck–Riemann–Roch theorem}
\(\mbox{ch}(f_{\mbox{!}}{\mathcal F}^\bullet)\mbox{td}(Y) = f_* (\mbox{ch}({\mathcal F}^\bullet) \mbox{td}(X))\). 
The smooth quasi-projective schemes $X$ and $Y$ in the theorem lack Wikipedia articles.
However, dedicated articles on\textit{ quasi-projective variet}y and \textit{smooth scheme} exist.
We proposed modeling this situation by creating the new Wikidata item \href{https://www.wikidata.org/wiki/Q85397895}{\emph{smooth quasi-projective scheme}}, which links to the existing articles as subclasses.
To create a clickable link from the Wikidata item to Wikipedia, we could create a new Wikipedia article on \textit{smooth quasi-projective scheme}.
Alternatively, we could add a new section on \textit{smooth quasi-projective scheme} to the article on \emph{quasi-projective variety} and create a redirect from the Wikidata item to the new section.

Aside from implementing the new feature, defining a decision-making process for the integration of math rendering features into Wikipedia was equally important.
For this purpose, we founded the \href{https://meta.wikimedia.org/wiki/Wikimedia_Community_User_Group_Math}{Wikimedia Community Group Math} as an international steering committee with authority to decide on future features of the math rendering component of Wikipedia.

\section{Conclusion \& Future Work}\label{sec.concl}
After working on Wikipedia's math support for several years, we have deployed the first feature that goes beyond improving the display of formulae. 
Realizing the feature became possible through the inauguration of the Wikimedia Community Group Math.

The new feature helps Wikipedia users to better understand the meaning of mathematical formulae by providing details on the elements of formulae. 
Because the new feature is available in all language editions of Wikipedia, all users benefit from the improvement. 
Rolling out the feature for all languages was important to us because using Wikipedia for more in-depth investigations is significantly more prevalent in languages other than English~\cite{LemmerichS0Z19}. 
Nevertheless, also in the English Wikipedia, fewer than one percent of the articles have a quality rating of good or higher~\cite{PiccardiCZ018}. 
Providing better tool support to editors can help in raising the quality of articles.
In that regard, our semantic enhancements of mathematical formulae will flank other semi-automated methods, such as recommending sections~\cite{PiccardiCZ018} and related articles~\cite{Schwarzer2016}. 

To stimulate the wide-spread adoption of semantic annotations for mathematical formulae, we are currently working on tools that support editors in creating the annotations.
With \texttt{AnnoMathTex} \cite{Scharpf2019b}, we are developing a tool that facilitates annotating mathematical formulae by providing a graphical user interface that includes machine learning assisted suggestions~\cite{moi} for annotations.
Moreover, we will integrate a field into the visual wikitext editor that will suggest Wikipedia authors to link the Wikidata id of a formula if the formula is in the Wikidata database.
Improved tool support will particularly enable smaller language editions of Wikipedia to benefit from the new feature because the annotations performed in any language will be available in all languages automatically.

  \paragraph{Acknowledgments}
  We thank the Wikimedia Foundation, particularly Marco Obrovac, for their continuous support. Our research was supported by the German Research Foundation \href{http://gepris.dfg.de/gepris/projekt/350192710}{DFG GI 1259/1}.
\printbibliography[keyword=primary]

@inproceedings{Schubotz2020,
  author    = {Schubotz, Moritz and Greiner-Petter, Andr\'{e} and Meuschke, Norman and Teschke, Olaf and Gipp, Bela},
  booktitle = {Proceedings of the ACM/IEEE Joint Conference on Digital Libraries in 2020 (JCDL '20), August 1--5, 2020, Virtual Event, China},
  date      = {2020},
  doi       = {10.1145/3383583.3398557},
  preprint  = {https://arxiv.org/pdf/2003.09417},
  title     = {Mathematical Formulae in Wikimedia Projects 2020},
}
\end{document}